# Resistance Drift in $Ge_2Sb_2Te_5$ Phase Change Memory Line Cells at Low Temperatures and Its Response to Photoexcitation

R. S. Khan[1], F. Dirisaglik[1,2], A. Gokirmak[1] and H. Silva[1]
[1]*Department of Electrical and Computer Engineering, University of Connecticut, Storrs, Connecticut 06269, USA*
[2]*Department of Electrical and Electronics Engineering, Eskisehir Osmangazi University, Eskisehir 26480, Turkey*

Resistance drift in phase change materials is characterized in amorphous phase change memory line-cells from 300 K to 125 K range and is observed to follow the previously reported power-law behavior with drift coefficients in the 0.07 to 0.11 range in dark. While these drift coefficients measured in dark are similar to commonly observed drift coefficients (~0.1) at and above room temperature, measurements under light show a significantly lower drift coefficient (0.05 under illumination versus 0.09 in dark at 150K). Periodic on/off switching of light shows sudden decrease/increase of resistance, attributed to photo-excited carriers, followed by a very slow response (~30 minutes at 150 K) attributed to contribution of charge traps. Continuation of the resistance drift at low temperatures and the observed photo-response suggest that resistance drift in amorphous phase change materials is predominantly an electronic process.

Phase change memory (PCM) has entered the main-stream non-volatile memory market due to its high-speed, endurance, and cost-effective scalability[1]. PCM utilizes the large contrast in conductivity between the crystalline (low-resistance) and amorphous (high-resistance) phase of chalcogenide materials such as $Ge_2Sb_2Te_5$ (GST) to store information[2]. PCM devices can be switched between these two phases using short electrical pulses[3] with endurance levels exceeding $10^{12}$ cycles[4]. The implementation of multi-bit-per-cell PCM is limited as resistance levels overlap with time due to the drift of the resistance of the amorphous phase with time, following a power-law[5]. Resistance drift has been reported to be a function of temperature[6–9], programmed resistance level[10,11], and read current[12]. There has been a number of theories explaining the origin of drift[5,13,14], however, the fundamental reason behind drift still needs to be fully understood.

Most reports in literature on resistance drift in PCM cells are on characterization studies at room temperature and above, where multiple processes may be occurring simultaneously. In this work, resistance drift in amorphous GST line cells is monitored in 300 K to 125 K temperature (T) range and the effect of light on the cell resistances is investigated.

The line-cells used in this work were fabricated using 90 nm technology with 250 nm thick bottom metal (Tungsten with Ti/TiN liner) contacts on 700 nm thermally grown $SiO_2$. GST was deposited using sputtering, patterned using optical lithography and reactive ion etching (RIE) and capped with 10 nm plasma enhanced chemical vapor deposition (PECVD) $Si_3N_4$ to prevent evaporation of Te during experiments (FIG. 1)[15]. The line cells used in these experiments were 50 nm in thickness with widths (W) varying from ~120 nm to ~140 nm and lengths (L) varying from ~390 nm to 500 nm. A schematic of the measurement setup is shown in FIG. 2. An Agilent 4156C semiconductor parameter analyzer and Tektronix AFG3102 arbitrary function generator were connected to a probe manipulator inside Janis cryogenic probe station through a relay controlled by an Arduino Mega 2560 microcontroller. A series resistor of 1 kΩ on the probe arm was used to limit the current during melting. The relay was not used for T ≤ 150K as the current going through the PCM cells becomes comparable to the leakage through the relay (~1pA). Pulse measurements were monitored using Tektronix DPO 4104 digital phosphor oscilloscope as shown in FIG. 2. The chamber was under high vacuum (~$10^{-5}$ Torr). Liquid nitrogen was used to cool the chamber and the chuck and a Lake Shore 336 temperature controller was used to control the temperature of the chuck.

The line-cells used in the experiment were annealed to 675

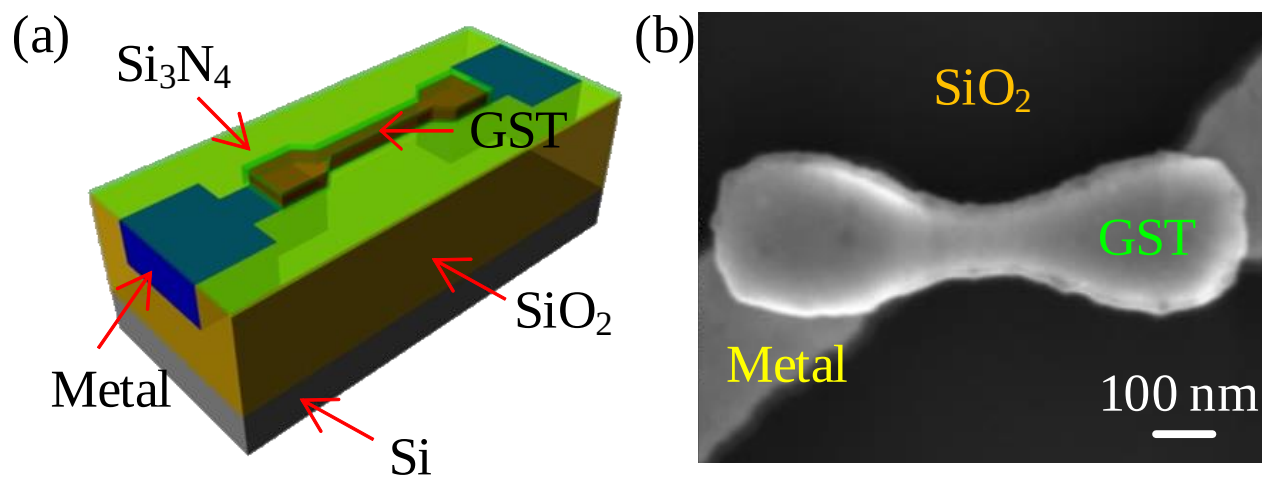

FIG. 1. Schematic (a) and scanning electron microscope (b) image of line cells used in the experiment.

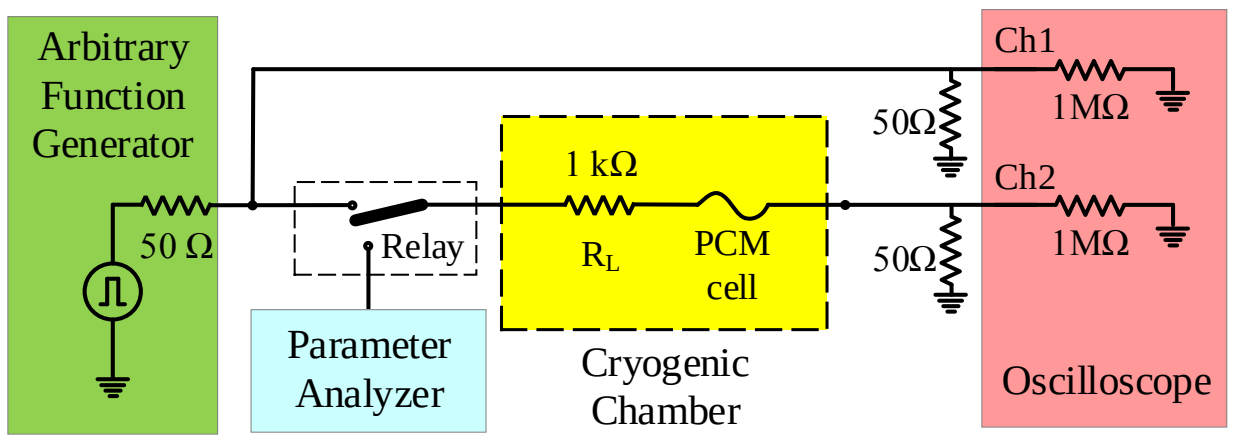

FIG. 2. Schematic of the measurement setup.

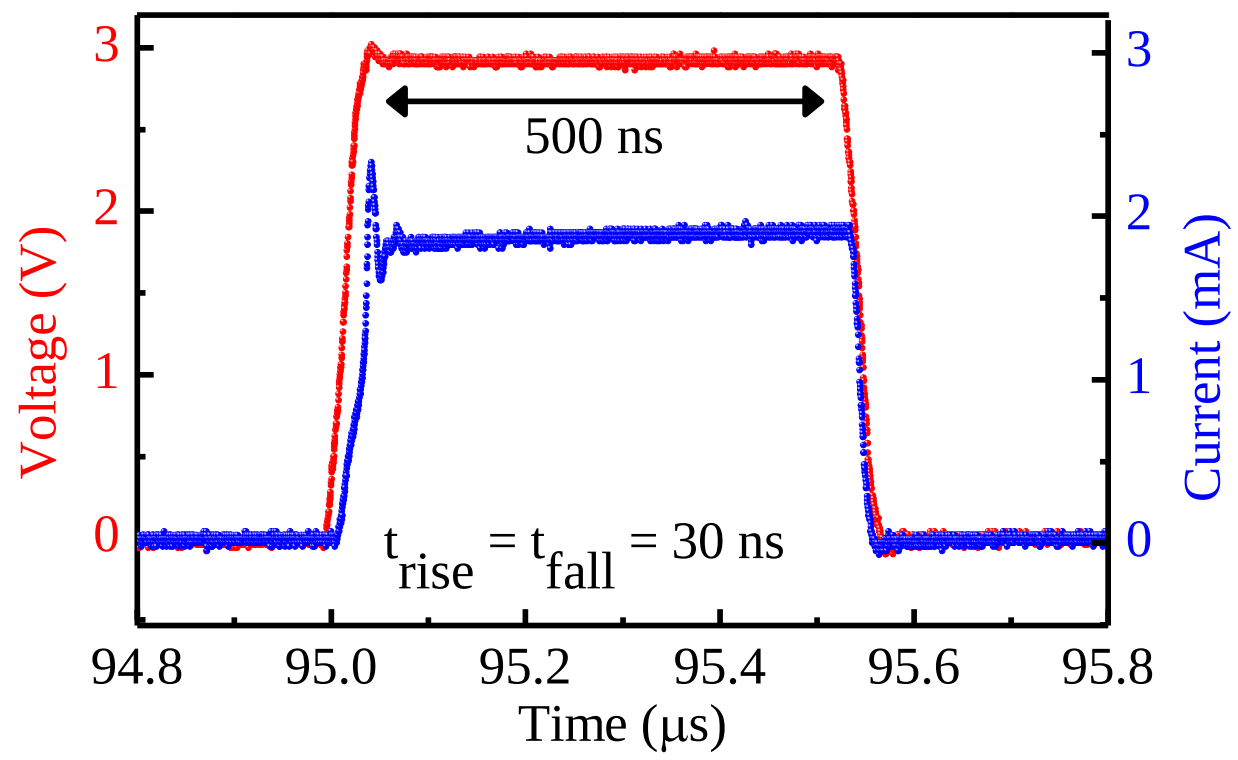

FIG. 3. Voltage used to amorphize the devices (red) and the resulting current (blue) waveforms as measured by the oscilloscope. Current was calculated using the voltage across the 50 Ω termination.

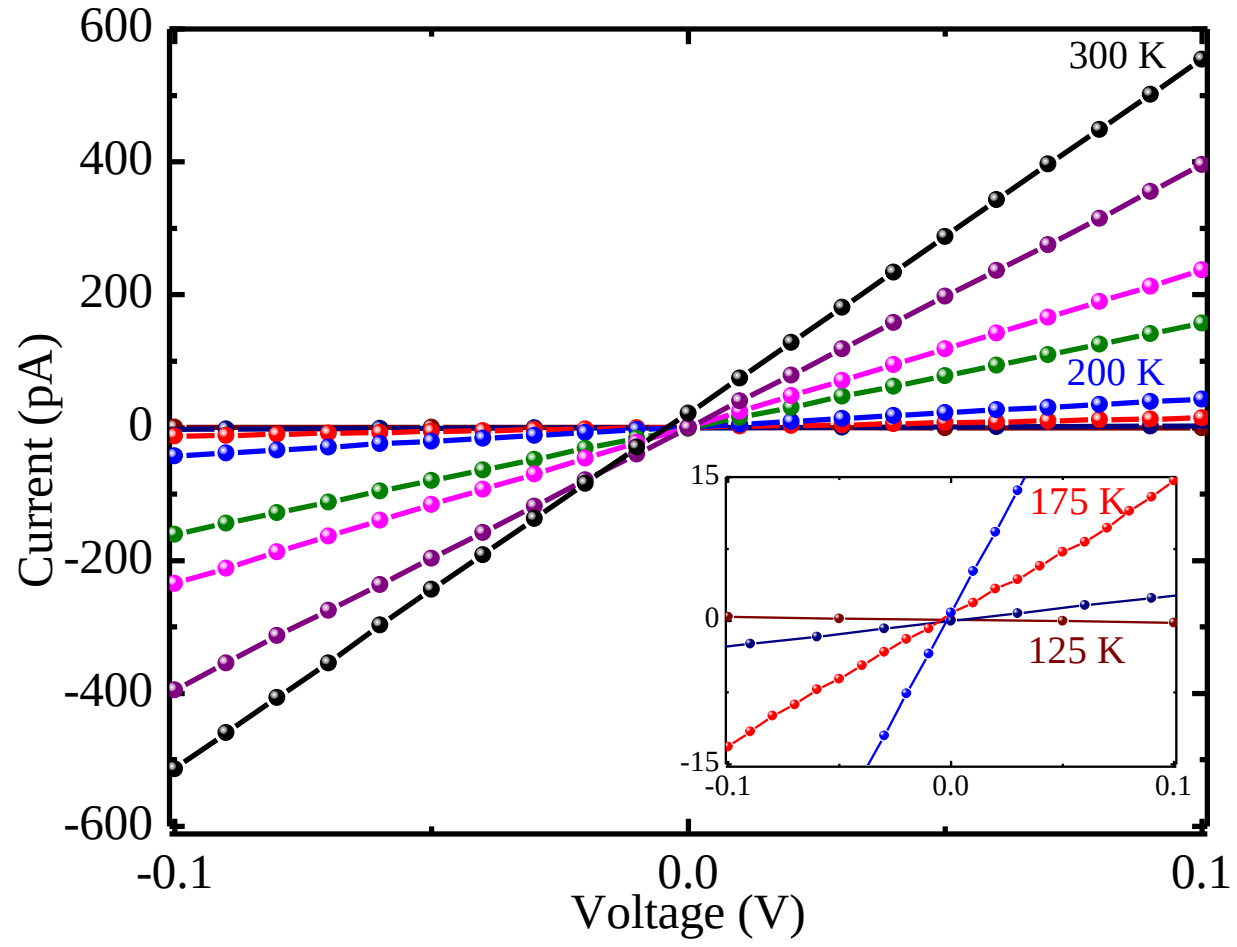

FIG. 4. Current versus voltage graphs for different temperatures in the 125 K to 300 K range. The inset zooms in to show 125 K, 150 K, 175 K and 200 K data.

K for 20 minutes under high vacuum to ensure that they were all in crystalline (HCP) phase prior to application of electrical pulses. A 500 ns pulse with 30 ns rise and fall times was used to amorphize each line-cell (FIG. 3). The cell resistances were measured using the parameter analyzer with low voltage sweeps to minimize disturbance of the state of the cells (FIG. 4). Multiple line-cells were amorphized at each temperature and the resistance of the cells was monitored from ~25 s after the pulse to ~$10^4$ s (with the exception of 125 K and 150 K, where the resistance was monitored starting from ~100 s after the pulse due to the delay associated with manual changes in the connections). The cell resistances were monitored at the temperature they were amorphized at.

The amorphized cell resistances were observed to increase with time following the commonly reported power-law[5]:

$$R = R_0 \left(\frac{t}{t_0}\right)^{\gamma}, \tag{1}$$

where R and $R_0$ are resistances at time t and $t_0$ respectively and γ is the drift coefficient. The drift coefficients can be extracted from the slope of the bi-logarithmic resistance (R) versus time (t) plots (FIG. 5). We see a slight increase in drift coefficient as a function of temperature, going from ~0.07 (125 K) to ~0.11 (300 K) (FIG. 6). Although resistance drift has often been attributed to structural relaxation, a thermally activated atomic rearrangement of the amorphous structure[13], the presence of resistance drift at cryogenic temperatures with similar drift coefficients as room temperature points to additional mechanisms contributing to this phenomenon.

Amorphous phase change materials are known to have a large number of charge traps in their band-gaps[16]. The charged traps perturb the local potential profile, and can change the local electric field, suppressing conduction[15]. If the capturing and emission of the charges at the traps play a significant role in resistance drift of amorphized cells, a response to photo-excitation is expected. In order to observe the effect of photo-excitation, a white LED (5 mm round top, driven at rated voltage 3.2 V and current 20 mA, color temperature 6000-9000 K, peak wavelength ~450 nm) was installed inside the cryogenic chamber, controlled by the

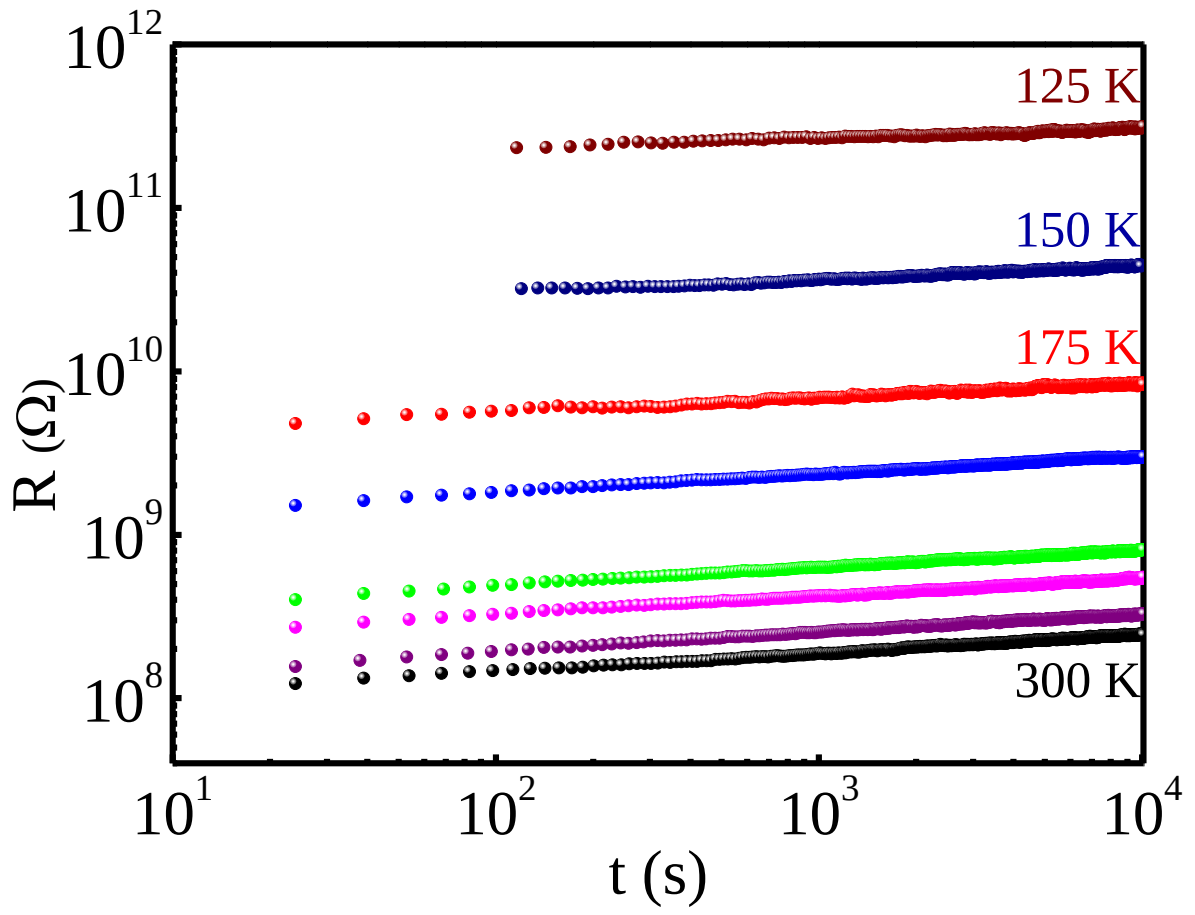

FIG. 5. Resistance (R) versus time (t) plots for line cells amorphized at temperatures from 125 K to 300 K at 25 K intervals. Measurements using a relay enable monitoring of the resistance in the 25 s -100 s period.

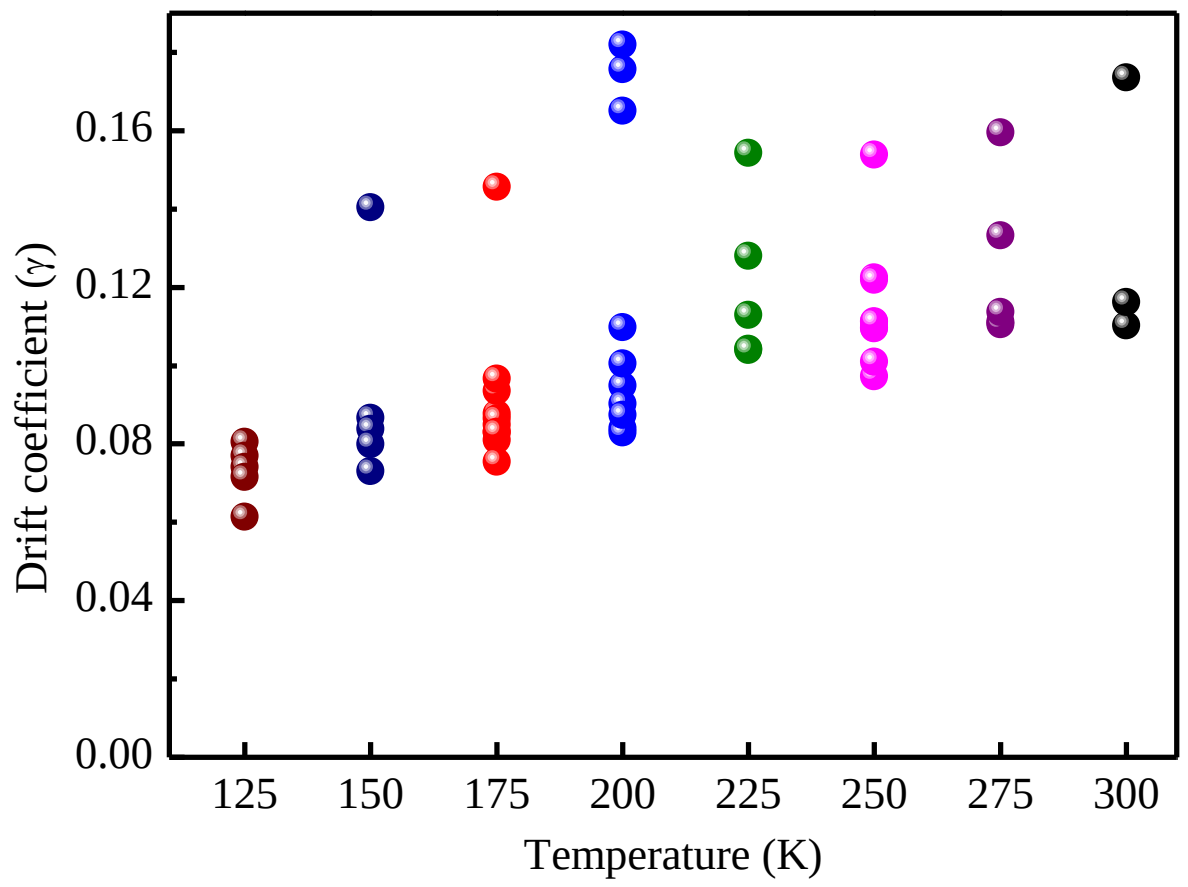

FIG. 6. Drift coefficients at different temperatures extracted from R-t plots. Each data point represents the drift coefficient of a different amorphized line cell of similar dimensions at respective temperature.

Arduino microcontroller. A low-power LED was chosen to minimize the thermal perturbation on the sample. The thermal relaxation timescales (~$10^{-7}$-$10^{-6}$ s)[6] are expected to be much less than the switching time-scale of the LED (~$10^{-1}$ s) and the measurement time-scale (~15 s).

During the LED on versus off experiments, two effects of light were observed: (i) electrical conductivity is higher under light, which is expected as amorphous GST (a-GST) is a semiconductor, (ii) line-cells amorphized and monitored under light showed significantly smaller drift than cells amorphized and monitored under dark (FIG. 7).

In a second set of experiments, the LED was periodically switched on and off, soon after amorphization (FIG. 8). A fast response followed by a gradual change in resistance was observed as the light was turned off and on. In the very long period measurements (30 minutes on and off durations), the resistance is observed to converge to a final value in ~10 minutes when the light is turned on and resistance is observed to converge to a resistance drift trajectory in ~10 minutes when the light is turned off (FIG. 8e).

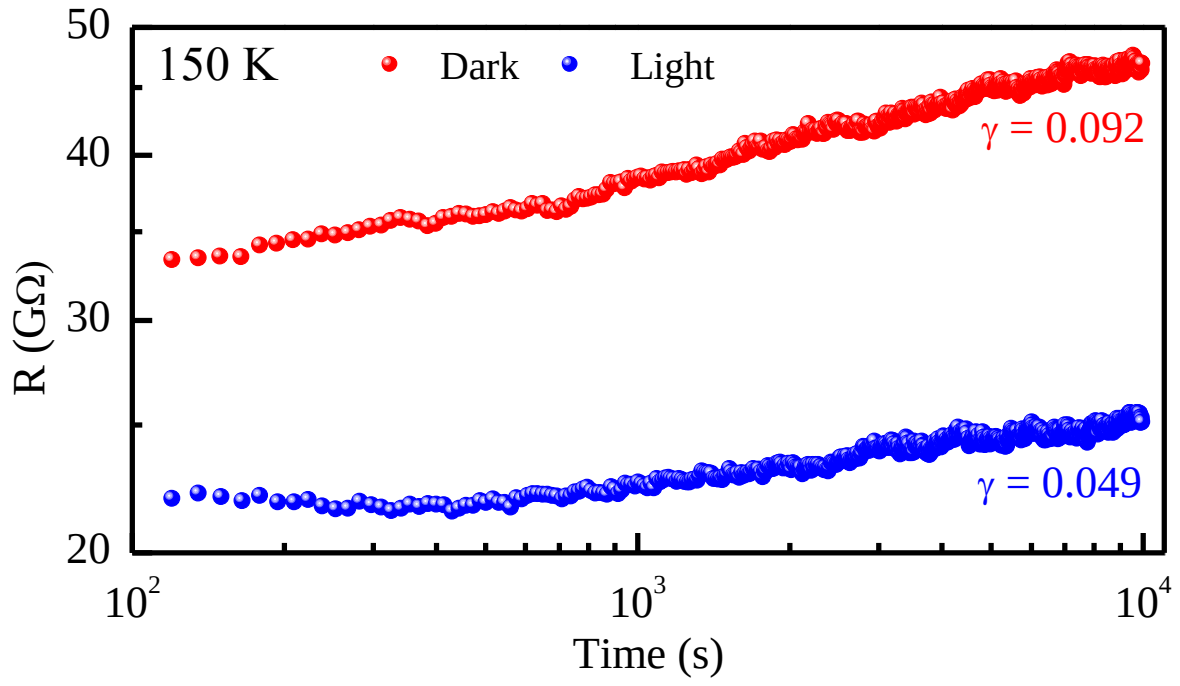

FIG. 7. Resistance versus time plots for two line cells amorphized and monitored under dark (red) and light (blue) at 150 K (a).

The fast changes in resistance can be attributed to photo-excited charge carriers. The long time-scale associated with these changes point to emptying of the traps when the light is turned on and filling of the traps when the light is turned off[17–19]. This extremely slow response suggests presence of slow trap-to-trap transitions similar to what is expected to give rise to phosphorescence[20] observed in high trap density material systems.

In summary, amorphized GST line cells display resistance drift in T = 300 K to 125 K, the full range of temperatures used in these experiments. Photo-excitation gives rise to a clearly observable change in resistance and lower drift coefficients, suggesting that photo-excited carriers have a significant contribution to conductance at low temperatures. Experiments with periodic exposure to light show a fast and a slow response. The time-scale of the slow response (~10 minutes at 150 K) suggest that very slow trap-to-trap charge exchanges have a significant impact on electrical conductivity and resistance can be stabilized under-light within each cycle. These results suggest that resistance drift in amorphous GST is predominantly an electronic process.

## Acknowledgement

Raihan Khan performed the experiments, analysis and writing of the manuscript supported by US NSF through grant # ECCS 1711626. The devices were fabricated at IBM T.J. Watson Research Center under a joint study agreement, by Faruk Dirisaglik supported by US DOE Office of Basic Energy Sciences (BES) and Turkish Educational Ministry. Ali Gokirmak and Helena Silva contributed to the design of experiments, analysis and writing of the manuscript. The authors are grateful to contributions of Adam Cywar of University of Connecticut, supported through US NSF Graduate Research Fellowship, and Chung Lam of IBM Research, for device fabrication. The authors would also like to thank Sadid Muneer and Nafisa Noor of University of Connecticut for valuable discussions.

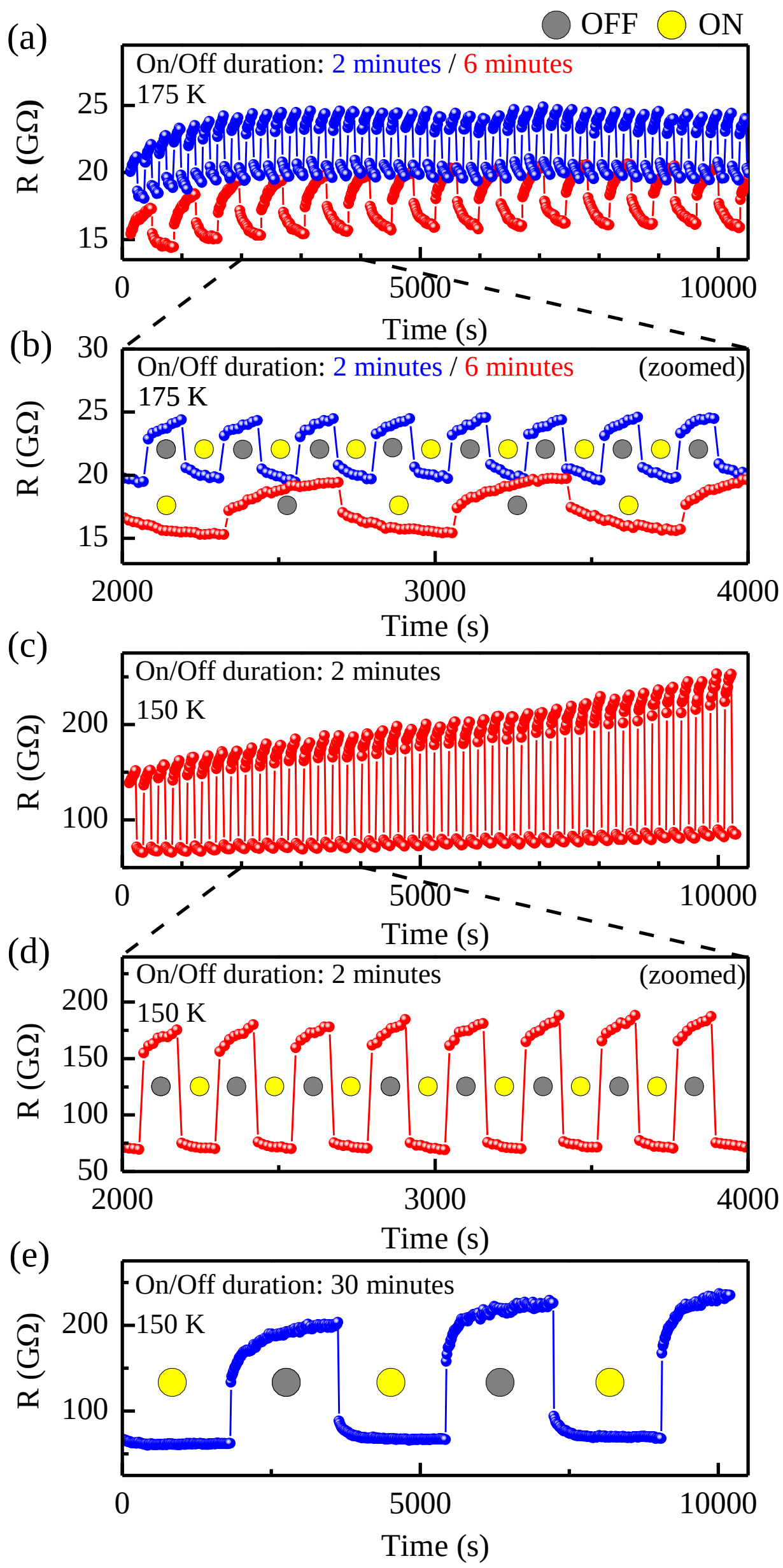

FIG. 8. Resistance versus time plots at 175 K (a, b) and 150 K (c, d, e) for different periods of photo-excitation using LED. The yellow (gray) circle indicates the LED being on (off) during that period. At both temperatures we observe a fast change in resistance followed by a slow response when the LED is switched on/off.